\documentclass[twocolumn,showpacs,preprintnumbers,amsmath,amssymb,floatfix]{revtex4}

\usepackage{graphicx}
\usepackage{dcolumn}
\usepackage{bm}

\begin{document}

%

\let\a=\alpha      \let\b=\beta       \let\c=\chi        \let\d=\delta
\let\e=\varepsilon \let\f=\varphi     \let\g=\gamma      \let\h=\eta
\let\k=\kappa      \let\l=\lambda     \let\m=\mu
\let\o=\omega      \let\r=\varrho     \let\s=\sigma
\let\t=\tau        \let\th=\vartheta  \let\y=\upsilon    \let\x=\xi
\let\z=\zeta       \let\io=\iota      \let\vp=\varpi     \let\ro=\rho
\let\ph=\phi       \let\ep=\epsilon   \let\te=\theta
\let\n=\nu
\let\D=\Delta   \let\F=\Phi    \let\G=\Gamma  \let\L=\Lambda
\let\O=\Omega   \let\P=\Pi     \let\Ps=\Psi   \let\Si=\Sigma
\let\Th=\Theta  \let\X=\Xi     \let\Y=\Upsilon

%

%

\def\cA{{\cal A}}                \def\cB{{\cal B}}
\def\cC{{\cal C}}                \def\cD{{\cal D}}
\def\cE{{\cal E}}                \def\cF{{\cal F}}
\def\cG{{\cal G}}                \def\cH{{\cal H}}
\def\cI{{\cal I}}                \def\cJ{{\cal J}}
\def\cK{{\cal K}}                \def\cL{{\cal L}}
\def\cM{{\cal M}}                \def\cN{{\cal N}}
\def\cO{{\cal O}}                \def\cP{{\cal P}}
\def\cQ{{\cal Q}}                \def\cR{{\cal R}}
\def\cS{{\cal S}}                \def\cT{{\cal T}}
\def\cU{{\cal U}}                \def\cV{{\cal V}}
\def\cW{{\cal W}}                \def\cX{{\cal X}}
\def\cY{{\cal Y}}                \def\cZ{{\cal Z}}

%

\newcommand{\Ns}{N\hspace{-4.7mm}\not\hspace{2.7mm}}
\newcommand{\qs}{q\hspace{-3.7mm}\not\hspace{3.4mm}}
\newcommand{\ps}{p\hspace{-3.3mm}\not\hspace{1.2mm}}
\newcommand{\ks}{k\hspace{-3.3mm}\not\hspace{1.2mm}}
\newcommand{\des}{\partial\hspace{-4.mm}\not\hspace{2.5mm}}
\newcommand{\desco}{D\hspace{-4mm}\not\hspace{2mm}}

\def\deltaakpi{\Delta {{A}}_{K\pi}}


\title{\boldmath Bounding
 $b\to s\mu^+\mu^-$  tensor operators from $B\to K^*(X_s) \gamma$}

\author{Namit Mahajan
}
\email{nmahajan@prl.res.in}
\affiliation{
 Theoretical Physics Division, Physical Research Laboratory, Navrangpura, Ahmedabad
380 009, India
}


\begin{abstract}
Tensor operators are often invoked as specific new physics operators beyond the standard model in an effort to explain anomalies in rare B-decays and CP asymmetries. Specifically, $b \to s \mu^+\mu^-$
tensor operators are invoked in the study of semi-leptonic decays, both inclusive and exclusive. In this note we use the data on $b \to s$ radiative decay modes and CP asymmetries to tightly constrain the tensor operators. It is found that constraints thus obtained are tighter than those from semi-leptonic modes. We also comment on $b \to ss\bar{s}$ tensor operators that help in explaining the $B \to \phi K^*$ polarization puzzle, and $b\to s\tau^+\tau^-$ operators with tensor structure.
\end{abstract}

\pacs{
 13.25.Hw, 
 13.20.-v, 
 12.60.-i 
 }
\maketitle


Absence of flavour changing neutral currents at the tree level within the standard model
(SM) makes them very sensitive to quantum corrections due to heavy particles in the
loops. Rare decays, both inclusive and exclusive, of the type $b \to s \gamma$, $b \to s g$
and $b \to s \ell^+\ell^-$ are among the most promising channels in our search for
possible new physics (NP) beyond SM. The study of CP violation,
and its origin, has been one of the main aims of the B-factories. Thanks to excellent
experimental precision reached at the B-factories, and also at CLEO and TeVatron,
we now have accurate measurements of barnching ratios and CP asymmteries for many
rare decay processes. LHCb has to be added to this list and has already begun to yield
competitive results even with small amount of data collected till now (see for example \cite{:2011zq}, \cite{Aaij:2011aj}, \cite{Koppenburg:2011eh}). The situation is expected
to improve drastically over the next few years \cite{Meadows:2011bk}. It is also not improbable that the first
glimpse of NP or the absence of it at the electroweak scale will be from rare B-decays rather than the direct searches at LHC.

Till date SM has turned out to be consistent with almost all the available experimental data, though
there are some anomalies or unexplained features that seem to call for physics beyond SM (see for example \cite{Asner:2010qj}). Experimental observations and measurements have established the dominance of Cabibbo-Kobayashi-Maskawa (CKM) phase as the prominent source of CP violation as far as the low energy sector is concerned.
Any attempt to infer hints of new physics (NP) need to ensure that we have quantitatively
exhausted all the possibilities within SM, including sub-leading effects and any other neglected
contributions based on some assumptions. Semi-leptonic and radiative decays of B-mesons offer a
unique opportunity to explore the possibility of NP, including new sources of CP violation beyond the
CKM phase. The inclusive decays, radiative and semileptonic, are theoretically more under control while the exclusive decays are
relatively easier experimentally. At present, the inclusive rate $BR(B\to X_s\gamma)$ and the exclusive rate
$BR(B\to K^{(*)}\ell^+\ell^-)$ and associated lepton forward-backward asymmetry provide the most stringent constraints on any new physics model, even in a
quite model independent manner. Latest experimental results indicate good agreement with SM expectations for
these modes but also leave a bit of a room for NP.

When going beyond SM, new and heavier particles (more massive than the electroweak scale) can bring in
totally new contributions not present in SM. An example could be left-right symmetric models which
naturally lead to operators in the low energy theory that have different chiral structure than SM (see for example \cite{Beall:1981ze}).
Other popular examples include supersymmetric theories which not only have different chiral structures for the
operators but also bring along completely new operators with naturally large coefficients or strengths \cite{Bertolini:1990if}.
The aim of the current and future experiments is to accurately measure all possible observables and thereby end up tightly constraining the possible structures/operators, and if possible revealing the specific type of NP present.

The effective Hamiltonian responsible for the semi-leptonic and radiative $b\to s$ transitions within SM is given by \cite{Buras:1998raa}
\begin{eqnarray}
{\mathcal{H}}_{eff}^{SM} &=& -\frac{4 G_F}{\sqrt{2}}V_{tb}^*V_{ts}\Big[\sum_{i=1}^{10} C_i Q_i + \Big\{C_{7\gamma}Q_{7\gamma} \nonumber \\
&& +  C_{8g}Q_{8g} + C_{9V}Q_{9V} + C_{10A}Q_{10A} \nonumber \\
&& +  C_SQ_S + C_PQ_P + \sum_X C_X'Q_X'
\Big\}\Big] \nonumber \\
&& +  V_{ub}^*V_{us}[...]+ H.C \label{heffective}
\end{eqnarray}
where $C$'s are the relevant Wilson coefficients while $Q$'s are four fermion
operators. Here, $Q_{1,2}$ are the current-current operators, while $Q_{3-6}$ and $Q_{7-10}$ are
the QCD penguin and electroweak (EW) penguin operators. Operators $Q_5$, $Q_6$, $Q_7$ and $Q_8$ have
$(V-A)\otimes (V+A)$ structure while all others have $(V-A)\otimes (V-A)$ structure. $Q_{7\gamma}$ and
$Q_{8g}$ are the electromagnetic and chromomagnetic dipole operators while $Q_{9V}$ and $Q_{10A}$ are
the vector and axial-vector semi-leptonic operators. $Q_S$ and $Q_P$ are the scalar and pseudoscalar semi-leptonic operators. The primed operators can be obtained from the unprimed ones by making the replacement $L\leftrightarrow R$.
The terms proportional to $V_{ub}^*V_{us}$ are usually neglected due to the smallness of $\vert V_{ub}\vert$.
For what concerns us here, the explicit form of some of the operators is:
\begin{eqnarray}
Q_1 &=& (\bar{c}_{L\beta}\gamma^{\mu}b_{L\alpha})(\bar{s}_{L\alpha}\gamma^{\mu}c_{L\beta})\, , \nonumber \\
Q_2 &=& (\bar{c}_{L\alpha}\gamma^{\mu}b_{L\alpha})(\bar{s}_{L\beta}\gamma^{\mu}c_{L\beta})\, , \nonumber \\
Q_{7\gamma} &=& \frac{e}{16\pi^2}m_b(\bar{s}_{\alpha}\sigma^{\mu\nu}R b_{\alpha})F_{\mu\nu}\, , \nonumber \\
Q_{8g} &=& \frac{g_s}{16\pi^2}m_b(\bar{s}_{\alpha}\sigma^{\mu\nu}R \frac{\lambda^A_{\alpha\beta}}{2} b_{\beta})G^A_{\mu\nu}\, , \nonumber \\
Q_{9V} &=& \frac{e^2}{16\pi^2}m_b(\bar{s}_{\alpha}\gamma^{\mu}Lb_{\alpha})(\bar{\ell}\gamma_{\mu}\ell)\, , \nonumber \\
Q_{10A} &=& \frac{e^2}{16\pi^2}m_b(\bar{s}_{\alpha}\gamma^{\mu}Lb_{\alpha})(\bar{\ell}\gamma_{\mu}\gamma_5\ell)\, , \nonumber \\
Q_S &=& \frac{e^2}{16\pi^2}m_b(\bar{s}_{\alpha}Rb_{\alpha})(\bar{\ell}\ell)\, , \nonumber \\
Q_P &=& \frac{e^2}{16\pi^2}m_b(\bar{s}_{\alpha}Rb_{\alpha})(\bar{\ell}\gamma_5\ell)
\label{opstructure}
\end{eqnarray}

The SM Wilson coefficients at scale $\mu=m_b$ (approximately) read:
\[
C_1\sim -0.3,\,\,\, C_2 \sim 1.14,\,\,\,  C_{3-6} \sim {\mathcal{O}}(10^{-2}),\]
\[ C_{7,8}\sim {\mathcal{O}}(10^{-4}),
\,\,\, C_9 \sim -1.28\alpha,\,\,\, C_{10}\sim 0.33\alpha \]
\[ C_{7\gamma} \sim -0.31,\,\,\,
C_{8g} \sim -0.15,\,\,\, C_{9V} \sim 4.2,\,\,\, C_{10A} \sim -4.1
\]
SM Wilson coefficients are real save for the small imaginary contributions due to $V_{ub}$, which we neglect here.
Due to the extreme smallness of $C_{S,P} \sim m_{\ell}m_b/m_W^2$ within SM, the corresponding operators are usually neglected. The primed Wilson coefficients typically read $C_X' \sim (m_s/m_b)C_X$, implying that they are expected to be suppressed and hence neglected. Specifically, within SM
\[
 C_{7\gamma}' \sim -0.006,\,\,\, C_{8g}' \sim -0.003,\,\,\, C_{9V}' = 0 = C_{10A}'
\]

The scalar and pseudo-scalar operators can have enhanced coefficients in many extensions of NP beyond SM, and can be cleanly probed in modes like $B_s\to \mu^+\mu^-$ \cite{Choudhury:1998ze}. In any extension of SM, either the Wislon coefficients of the operators already present get new non-negligible
contributions or there are new operators induced in the low energy theory or both.
 In many analyses of semi-leptonic decays, tensor operators are considered \cite{Fukae:1998qy}, \cite{Alok:2010zd}. Tensor operators are also invoked to explain the polarization puzzle
in $B \to \phi K^*$ since they have the capability to significantly enhance the transverse polarization fraction and hence explain the experimental data \cite{Das:2004hq}. We consider the following tensor operators with the scale of new physics denoted by $\Lambda$
\begin{eqnarray}
Q^T_{LL} &=& (\bar{s}_{\alpha}\sigma^{\mu\nu}Lb_{\alpha})(\bar{f}\sigma_{\mu\nu}L f)\, , \nonumber \\
Q^T_{LL} &=& (\bar{s}_{\alpha}\sigma^{\mu\nu}Lb_{\alpha})(\bar{f}\sigma_{\mu\nu}R f)\, , \nonumber \\
Q^T_{LL} &=& (\bar{s}_{\alpha}\sigma^{\mu\nu}Rb_{\alpha})(\bar{f}\sigma_{\mu\nu}L f)\, , \nonumber \\
Q^T_{LL} &=& (\bar{s}_{\alpha}\sigma^{\mu\nu}Rb_{\alpha})(\bar{f}\sigma_{\mu\nu}R f)
\label{tensorops}
\end{eqnarray}
such that the additional terms in the effective Hamiltonian read
\begin{eqnarray}
{\mathcal{H}}_{eff}^{NP} &=& -\frac{1}{\Lambda^2}\sum_{AB}{\cal{C}}^T_{AB}Q^T_{AB} \nonumber \\
&=& -\frac{4 G_F}{\sqrt{2}}V_{tb}^*V_{ts}\sum_{AB}C^T_{AB}Q^T_{AB}
\end{eqnarray}
In the above equations, $f$ refers to any light charged fermion. We shall be specifically interested in $f=\mu$.

The Wilson coefficients are in general complex quantities but to simplify the analysis we assume
them to be real here. This then broadly refers to Minimal Flavour Violation scenario where it is assumed that
CKM is the only source of CP violation (see for example \cite{Buras:2003jf}). This should only be taken as a simplifying assumption when
trying to make model independent statements. Else, we should write all possible relevant operators and allow for
complex coefficients. Then a detailed fit to the data would yield the best fit values for the coefficients. This however requires a very large data set and a complicated analysis. In the meantime, one could just focus on
a smaller sub-set of operators and try to constrain them in a somewhat independent fashion. This is the typical approach that is followed generally and we also adhere to that for the present study.
Given the new tensor operators, the next task would be to study their effect on various processes. The obvious ones $f=\mu$ are the semileptonic channels like $B \to K^{(*)}(X_s)\mu^+\mu^-$. By comparing the theoretical branching ratios and other observables like forward-backward, CP and any other possible asymmetries with the experimentally available data, constraints on the coefficients of the new operators are obtained. For example, the rate of the inclusive semileptonic process $B\to X_s\mu^+\mu^-$ leads to a relation $\vert C_T\vert^2 + 4\vert C_{TE}\vert^2 <1$ \cite{Alok:2010zd}, where $C_T$ and $C_{TE}$ are the coefficients of the following two operators that are usually considered in the literature:
\begin{eqnarray}
O_T &=& (\bar{s}\sigma^{\mu\nu}b)(\bar{\mu}\sigma_{\mu\nu}\mu) \nonumber \\
O_{TE} &=& i(\bar{s}\sigma^{\mu\nu}b)(\bar{\mu}\sigma^{\alpha\beta}\mu)\epsilon_{\alpha\beta\mu\nu}
\label{tensormumu}
\end{eqnarray}
It is clear that the above two operators in Eq.(\ref{tensormumu})when added in suitable combinations are equivalent to the four tensor operators listed in Eq.(\ref{tensorops}). There is however one small but potentially crucial difference. The Wilson coefficients of operators generated via suitable linear combinations of the operators $O_T$ and $O_{TE}$ are not all different, and therefore much more tightly constrained, although there is a priori no reason for some of the coefficients to be equal.

We now study the effect of the tensor operators on $b \to s\gamma$ processes. It is clear that due to the dipole structure of the operators involved, these operators will directly contribute to $b \to s\gamma$. Fig. 1 shows the Feynman diagrams for operator insertions leading to new contributions to the process. For the present case, we study the effects when $f = \mu$. Thus only the left diagram contributes (other possible diagrams where the photon is attached to the external quark lines are not shown). To evaluate the effect of all the four tensor operators listed in Eq.(\ref{tensorops}), we consider the following general structure
\begin{equation}
 Q^T = (\bar{s}_{\alpha}\sigma^{\mu\nu}\frac{1}{2}(1 - a\gamma^5){2}b_{\alpha})(\bar{f}\sigma_{\mu\nu}\frac{1}{2}(1 - a'\gamma^5)f)
\label{genop}
\end{equation}
with $a, a' = \pm 1$.
\begin{figure}[ht!]
\vskip -1.2cm
\hskip -2.35cm
\hbox{\hspace{0.03cm}
\hbox{\includegraphics[scale=0.75]{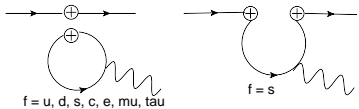}}
}
\caption{Feynman diagrams (drawn using the package JaxoDraw \cite{Binosi:2003yf}) for generating $b\to s\gamma$ via the operator insertions (crossed circles denote the operator insertions). The right hand diagram gives the second insertion possible when the light fermion $f$ is the strange quark.
 }
 \label{fig1}
\end{figure}

On evaluating the diagram with the operator in Eq.(\ref{genop}) one finds that the chirality factors involved yield a non-zero contribution only if $a=a'$. This simply implies that only $Q^T_{LL}$ and $Q^T_{RR}$ contribute and can thus be bounded. The other two operators are totally unconstrained from the present analysis. This is precisely the potentially important difference between the basis considered in Eq.(\ref{tensorops}) and the one usually employed in the study of semi-leptonic decays, ie the one in Eq.(\ref{tensormumu}). On evaluating the loop diagram, one finds for both $a=a'=+1$ and $a=a'=-1$, the same factor
\[
F_{Loop} = 16 Q_f \frac{m_f}{m_b}\ln\left(\frac{m_f^2}{\mu_R^2}\right)
\]
where $Q_f$, $m_f$ and $\mu_R$ are the light fermion charge, mass and renormalization scale respectively. We set $\mu_R = m_b = 4.8$ GeV. The extra explicit factor of $m_b$ in the above expression has been introduced for convenience such that the new contribution finally takes the familiar form of $Q_{7\gamma}^{(')}$. Denoting by $\Delta C_{7\gamma}$ and $\Delta C_{7\gamma}'$ the contributions to respective Wilson coefficients due to the new physics effects, one has
\begin{equation}
\Delta C_{7\gamma}^{(')} = F_{Loop}C^T_{RR(LL)}
\label{c7floop}
\end{equation}
For the case of muon, $F_{Loop}^{\mu} \sim 2.5$. For the case of strange quark, the $F_{Loop}^s \sim 0.8$. However, for the tau lepton or charm quark, $F_{Loop}^{c,\tau} \sim \mathcal{O}(10)$. For the present analysis, we set $F_{Loop}^{\mu} = 2$. It may be worthwhile to mention that the Wilson coefficients for the tensor operators at scales $m_b$ and $m_W$ are related by \cite{Bobeth:2011st}
\begin{equation}
C^T_{AB}(m_b) = \left(\frac{\alpha_s(m_W)}{\alpha_s(m_b)}\right)^{4/(3\beta_s)}C^T_{AB}(m_W) \sim \mathcal{O}(1)
\end{equation}
where $\beta_s = 11 - 2N_f^{active}/3$. This then implies that as a first approximation, the changes due to running may be neglected.
One therefore has the following at the scale $m_b$:
\begin{equation}
C_{7\gamma} \to C_{7\gamma} + \Delta C_{7\gamma}; \hskip 0.5cm C_{7\gamma}' \to C_{7\gamma}' + \Delta C_{7\gamma}'
\end{equation}
We note in passing that if the light fermion $f$ is a quark, then the same set of operators will also contribute to the chromomagnetic dipole operators ie to $\Delta C_{8g}^{(')}$. The answers can be easily read off after making appropriate changes in  $\Delta C_{7\gamma}^{(')}$. At the scale $m_b$, the Wilson coefficients mix and read \cite{Kagan:1998bh}
\begin{eqnarray}
C_{7\gamma}(m_b) &\sim& -0.31 + 0.67C_{7\gamma}^{NP}(m_W) + 0.09C_{8g}^{NP}(m_W) \nonumber \\
C_{8g}(m_b) &\sim& -0.15 + 0.70C_{8g}^{NP}(m_W)
\end{eqnarray}
It is the extra contribution which is labeled $\Delta C_{7\gamma,8g}$ and when translating into constraints on the coefficients of specific tensor operators, these relations could be inverted and the constraints read off.

The branching ratio and time dependent CP asymmetry have been measured for $B\to K^*\gamma$. Also available are the very precise measurement of the inclusive branching fraction $BR(B\to X_s\gamma)$. The direct CP asymmetry in the inclusive radiative mode can also be considered. The experimental situation is summarised in Table 1.

\begin{table}
\begin{center}
\begin{tabular}{|l|l|}
\hline
Observable & HFAG average \cite{Asner:2010qj}\\ \hline
$BR(B\to K^*\gamma)$ & $(42.7\pm 1.8) \times 10^{-6}$ \\
$BR(B\to X_s\gamma)$ & $(3.55\pm 0.26) \times 10^{-4}$ \\
$A_{CP}(b\to s\gamma)$ & $(-0.012\pm 0.028) \%$ \\
$S_{K^*\gamma}$ & $-0.16 \pm 0.22$\\ \hline
\end{tabular}
\caption{HFAG average values for observables in $b\to s\gamma$ system}
\end{center}
\end{table}

\begin{figure}[ht!]
\hbox{\hspace{0.03cm}
\hbox{\includegraphics[scale=0.615]{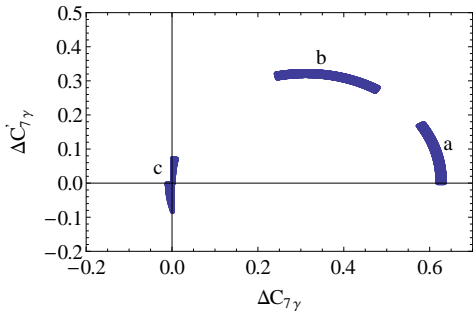}}
}
\caption{Scatter plot of allowed region in $\Delta C_{7\gamma}$-$\Delta C_{7\gamma}'$ plane. Both the parameters are varied from -1 to 1.
 }
 \label{fig2}
\end{figure}

We consider the exclusive mode first. The decay rate (or the branching fraction) reads \cite{Ali:2001ez}
\begin{eqnarray}
BR(B \to K^*\gamma) &=& (V_{tb}^*V_{ts})^2 \frac{G_F^2m_b^2m_B^3}{2^5\pi^4}\left(1-\frac{m_{K^*}^2}{m_B^2}\right)^3\nonumber \\
&& \times \vert T_1(0)\vert^2(\vert C_{7\gamma}\vert^2 + \vert C_{7\gamma}'\vert^2)
\end{eqnarray}
where $T_1$ is the form factor at $q^2=0$, while the time dependent CP asymmetry reads \cite{Atwood:1997zr}
\begin{equation}
S_{K^*\gamma} \simeq \frac{2}{\vert C_{7\gamma}\vert^2 + \vert C_{7\gamma}'\vert^2} Im(e^{-i\phi_d}C_{7\gamma}C_{7\gamma}')
\end{equation}
In the above equation, $\phi_d$ describes the mixing in the $B_d$ sector, ie $\sin(\phi_d) = S_{\psi K_S}$, time dependent mixing induced CP asymmetry. We employ the experimentally measured value $S_{\psi K_S} = 0.67\pm 0.02$ \cite{Asner:2010qj} in the numerical analysis. We consider $1\sigma$ and $2\sigma$ ranges for the branching ratio and mixing induced CP asymmetry respectively. The reason for choosing $2\sigma$ range for $S_{K^*\gamma}$ is to include possible error due to $S_{\psi K_S}$, and we then employ the central value in the analysis. Fig. 2 shows the constraints in $\Delta C_{7\gamma}$-$\Delta C_{7\gamma}'$ plane.
The constraints on $C^T_{LL,RR}$ are readily obtained from Eq.(\ref{c7floop}). From Fig. \ref{fig2} it is clear that demanding that $BR(B \to K^*\gamma)$ and $S_{K^*\gamma}$ are within the experimental ranges yields very tight constraints on the Wilson coefficients, more stringent than the maximally allowed ones from the semi-leptonic processes. As an example and a check, we looked at the representative values of $C_T$ and $C_{TE}$ employed in \cite{Alok:2010zd} and check whether they yield consistent values for both $BR(B \to K^*\gamma)$ and $S_{K^*\gamma}$. We find that for most of the representative pairs $(C_T, C_{TE})$ either or both the observables fail to fall within the experimentally allowed ranges. At this point it is rather important to clearly mention that the values employed in \cite{Alok:2010zd} are the ones that give maximal deviation from SM expectations for the observables studied. However, smaller values are also consistent with their analysis \cite{amol}. In no way this invalidates the analysis in \cite{Alok:2010zd} but the main point of this exercise was to illustrate that once the tensor operators are appropriately contracted in order to obtain additional contributions to $C_{7\gamma}$ and $C_{7\gamma}'$, only a restricted region of parameter space survives. This therefore shows the power and importance of combining
the constraints from a direct analysis like semi-leptonic modes and indirect ones like radiative modes.

Fig. \ref{fig2} has three regions which are allowed once constraints from $BR(B \to K^*\gamma)$ and $S_{K^*\gamma}$ are included. Region 'c' corresponds to small deviations from the SM values and hence is kind of expected. Bulk of the region 'a' corresponds to the case where the sign of $C_{7\gamma}$ gets flipped and $C_{7\gamma}'$ is not too large. This region naively speaking is strongly disfavoured from $B\to X_s\ell^+\ell^-$ rate. However, it is important to carefully check whether with $C_{7\gamma}'$ also present this result still holds or not. Region 'b' is the most interesting region since both the coefficients are non-negligible and of similar magnitude. A complete and consistent study will involve considering the effect of $C_{7\gamma}'$ (and other relevant chirality flipped operators) on semi-leptonic modes in conjunction with the radiative modes discussed here. This is beyond the scope of the present work and will be dealt elsewhere.

From Eq.(\ref{c7floop}) it is clear that if $f=\tau$, due to large $F_{Loop}^{\tau}$, the corresponding Wilson coefficients $C^{T,\tau}_{LL,RR}$ will be smaller. When translated in terms of ${\mathcal{C}}^T$, the values obtained are consistent with those obtained in \cite{Bobeth:2011st}.

We now consider the tensor operators when $f=s$. Such operators are invoked in order to explain the polarization puzzle in $B\to \phi K^*$ modes. Authors of \cite{Das:2004hq} study a host of observables available in $B\to \phi K^*$ modes and obtain the following best fit values: $C^T_{LL}(f=s) \sim 2\times 10^{-4}e^{i\phi_{LL}}e^{i\delta_{LL}}$ with $\phi_{LL}=-0.12, \hskip 0.2cm \delta_{LL}=1.15$
and $C^T_{RR}(f=s) \sim 1.7\times 10^{-4}e^{i\phi_{RR}}e^{i\delta_{RR}}$
 with $\phi_{RR}=0.14, \hskip 0.2cm \delta_{RR}=2.36$
where $\phi$'s and $\delta$'s are the weak and strong phases expressed in radians. We use these and find that they yield consistent values for both $BR(B \to K^*\gamma)$ and $S_{K^*\gamma}$.

We have explicitly checked that the inclusive $b\to s\gamma$ rate also yields similar constraints. Another powerful observable is the direct CP asymmetry in $B\to X_s\gamma$. Following \cite{Kagan:1998bh}, for a cut on photon energy, $E_{\gamma}>(1-\delta_{\gamma})E_{\gamma}^{max}$ with $\delta_{\gamma}=0.3$
\begin{eqnarray}
A_{CP}(b\to s\gamma) &\sim& \frac{0.01}{(\vert C_{7\gamma}\vert^2 + \vert C_{7\gamma}'\vert^2)}[1.17Im(C_2C_{7\gamma}^*) \nonumber \\
 &&- 9.51Im(C_{8g}C_{7\gamma}^* + C_{8g}C_{7\gamma}^{'*}) \nonumber \\
&& + 0.12Im(C_2C_{8g}^*) -9.40Im(\epsilon_sC_2 \\
&& \times (C_{7\gamma}^* - 0.0138C_{8g}^* + C_{7\gamma}^{'*} - 0.0138C_{8g}^{'*}))] \nonumber
\end{eqnarray}
where $\epsilon_s = V_{ub}^*V_{us}/(V_{tb}^*V_{ts})$.

We have checked that our results are consistent with $A_{CP}(b\to s\gamma)$, and so are the values for Wilson coefficients obtained by \cite{Das:2004hq}. It is interesting to notice that seemingly small Wilson coefficients as in \cite{Das:2004hq} which as expected would be consistent with the constraints from radiative modes still are able to explain the polarization puzzle in $B\to \phi K^*$. Very recently, a similar situation has arisen in $B_s\to K^{*0}\bar{K}^{*0}$ where again the longitudinal fraction is found to be much lower than the expectations \cite{Aaij:2011aj}. This is a $b\to s\bar{d}d$ penguin dominated mode and is expected to be an important channel in search of new physics. It would be interesting to see if similar tensor operators can explain the polarization puzzle and remain consistent with the constraints from radiative modes. We leave this for a separate study. It is also noteworthy that to explain the polarization puzzle other type of operators are also considered, eg right handed currents \cite{Kagan:2004ia}, 
(pseudo-)scalar operators \cite{Chen:2005mka}. We would like to emphasize that any operator of the form $(\bar{s}\Gamma b)(\bar{q}\Gamma'q)$, where $\Gamma,\,\Gamma'$ are Dirac structures would in principle generate new contributions to dipole operators and one should explictly check whether they pass the simple tests discussed above.

In this note we have studied the impact of tensor operators corresponding to physics beyond SM, that are invoked in the study of semi-leptonic decays $b\to s\ell^+\ell^-$ and $b\to s\bar{s}s$ to explain polarization puzzle in $B\to \phi K^*$ modes, on radiative modes $b\to s\gamma$ and CP asymmetries. We have shown that two of the tensor operators with chirality $LL$ and $RR$ can be stringently constrained. The other two operators with chiral structure $LR$ and $RL$ do not contribute to the radiative mode and therefore are left unconstrained from the present analysis. We have also eluded to a potential difference between the case when all the Wilson coefficients for these operators are taken as free parameters and the case when due to specific choice of the operators some of the coefficients are equal to each other. This according to us may be over restrictive. We have found that the tensor operators end up generating new contributions to dipole operators with both the chiralities. The Wilson coefficients have been assumed to be real but extension to complex coefficients is straightforward. It is known that complex coefficients yield a far more richer phenomenology (see for example \cite{Hovhannisyan:2007pb} for effects of complex coefficients on $B\to K^*\ell^+\ell^-$) and in fact presence of new phases may fake the effects naively expected due to the presence of new class of operators.
Also, if the operator under consideration has the light fermion as a quark, then a similar contribution is generated for the chromomagnetic dipole operators and therefore would contribute non-trivially to $A_{CP}(b\to s\gamma)$ and $BR(b\to sg)$. An enhanced $b\to sg$ rate may help in explaining some of the issues in rare B-decays like charm counting, large $B\to K\eta'$ rate and semi-leptonic branching ratio (see for example \cite{Kagan:1997qna}).
Further, the presence of chirality flipped dipole operator calls for a separate analysis including semi-leptonic decays. We leave all these for a future study. In conclusion, we have shown that rather tight constraints can be obtained on the tensor operators by studying their impact on radiative modes. With further improvement in experimental values, we hope that it may be possible to pin down the allowed range of the strength of these operators within very narrow bands. Similar remarks would apply to any other operator (any Dirac and chiral structure) with the light fermion being a quark and in many cases it may be plausible that some of the operators get tightly constrained or almost (practically) ruled out.

\vskip 0.3cm \noindent 

\end{document}